\RequirePackage{snapshot}
\documentclass[conference]{IEEEtran}
\usepackage{authblk}

\usepackage{amsfonts,amsmath,amsthm,amssymb,dsfont}

\ifCLASSINFOpdf 
\usepackage[pdftex]{graphicx} 
\usepackage[pdftex,colorlinks, 
           bookmarks=true,%
           pdftitle=Optimizing\ spatial\ throughput\ in\ device-to-device\ networks,%
           pdfauthor=B.\ Blaszczyszyn%
           \ &\ H.\ P.\ Keeler
           \ &\ P.\ Muehlethaler ,]{hyperref}  
\usepackage[numbers,sort&compress]{natbib} 
\usepackage{hypernat} 
\hypersetup{
   bookmarksnumbered,
   pdfstartview={FitH},
   citecolor={blue},
   linkcolor={red},
   urlcolor=[rgb]{0,0.55,0},
   pdfpagemode={UseOutlines}}
\else

 \usepackage[dvips]{graphicx} 
 \usepackage[numbers,sort&compress]{natbib} 
\fi 
 
\usepackage{epstopdf}

\newcommand{\E}{\mathbf{E}}

\newcommand{\calL}{\mathcal{L}}
\newcommand{\calP}{\mathcal{P}}

\newcommand{\Prob}{\mathbf{P}}

\newcommand{\R}{\mathbb{R}}

\newcommand{\Lap}{\mathcal{L}}

\newcommand{\SIR}{\text{SIR}}

\newcommand{\pathB}{{\ell}}
\newcommand{\pathD}{{\ell}}
\newcommand{\betaB}{{\beta}}
\newcommand{\betaD}{{\beta}}
\newcommand{\kappaB}{\kappa}
\newcommand{\kappaD}{\kappa}
\newcommand{\lambdaB}{{\lambda_{\textrm{B}}}}
\newcommand{\lambdaD}{{\lambda_{\textrm{D}}}}
\newcommand{\aB}{{a_{\textrm{B}}}}
\newcommand{\aD}{{a_{\textrm{D}}}}
\newcommand{\aI}{{a_{\textrm{I}}}}
\newcommand{\threshold}{\tau}
\newcommand{\thresholdB}{{\threshold_{\textrm{B}}}}
\newcommand{\thresholdD}{{\threshold_{\textrm{D}}}}

\newcommand{\throughput}{D}
\newcommand{\calPB}{{\calP_{\textrm{B}}}}
\newcommand{\calPD}{{\calP_{\textrm{D}}}}

\newcommand{\IB}{{I_{\textrm{B}}}}
\newcommand{\ID}{{I_{\textrm{D}}}}

\newcommand{\PB}{{P_{\textrm{B}}}}
\newcommand{\PD}{{P_{\textrm{D}}}}

\def\fade{{F}} 
\def\rayleigh{{E}} 

\def\fadeD{{\rayleigh}} 

\newtheorem{thm}{Theorem}
\newenvironment{theorem}{\bf\begin{thm}\rm\em}{\end{thm}} 
\newtheorem{cor}[thm]{Corollary}

\newtheorem{lem}[thm]{Lemma}
\newenvironment{lemma}{\bf\begin{lem}\rm\em}{\end{lem}} 
\newtheorem{prop}[thm]{Proposition}
\newenvironment{proposition}{\bf\begin{prop}\rm\em}{\end{prop}} 
\newtheorem{rem}[thm]{Remark}
\newenvironment{remark}{\bf\begin{rem}\rm}{\end{rem}} 

\title{
Optimizing spatial throughput in device-to-device networks
}

\author[1]{B. B{\l}aszczyszyn~\thanks{Bartek.Blaszczyszyn@ens.f}}
\author[2]{H. P. Keeler~\thanks{keeler@wias-berlin.de}}
\author[3]{P. M{\"u}hlethaler~\thanks{Paul.Muhlethaler@inria.fr}}
\affil[1]{Inria/ENS, Paris, France}
\affil[2]{Weierstrass Institute, 10117 Berlin, Germany}
\affil[3]{Inria,  Paris, France}

\begin{document}


\maketitle

\begin{abstract}
Results are presented for optimizing device-to-device  communications in cellular networks, while maintaining spectral efficiency of the base-station-to-device downlink channel. We build upon established and tested stochastic geometry models of signal-to-interference ratio in wireless networks based on the Poisson point process, which incorporate random propagation effects such as fading and shadowing. A key result is a simple formula, allowing one to optimize the device-to-device spatial throughput by suitably adjusting the proportion of active devices.  
These results can lead to further investigation as they can be immediately applied to more sophisticated models such as  studying multi-tier network models to address coverage in closed access networks.  
\end{abstract}


\section{Introduction}
Device-to-device  networks are available and emerging technologies that allow direct communication between devices in cellular phone networks, resulting in the need for new algorithms and methods in network resource management~\cite{doppler2009device,fodor2012design,bangerter2014networks}. 
Under a network model, we propose an optimization problem for device spatial throughput and then we present its simple solution, which network operators may use to better optimize device-to-device communication in cellular networks. Our approach is based on the information-theoretic concept of signal-to-interference ratio (SIR) in the downlink channel, as well as established and recent results from stochastic geometry models of wireless networks~\cite{sinrmoments}. The model essentially consists of combining a bi-polar model for devices, originally developed for  mobile ad hoc wireless networks, with a well-established SIR model for  cellular networks~\cite{andrews2011tractable}, with both models being based on the Poisson (point) process.  

Researchers have developed wireless network models based on the Poisson process, which serves as the foundation for the clear majority of stochastic geometry network models. 
Naturally, real-life cellular networks may not statistically resemble actual realizations of a Poisson process. But fortunately this issue can often be circumvented by rigorous mathematical results, which show that to any single observer the network will appear more Poisson-like in terms of received signal strengths, provided there are sufficient random propagation effects such as multi-path fading or randomly varying signal strengths~\cite{hextopoi-journal,keeler2014wireless}. In other words, a non-Poisson network can appear more Poisson due to the randomness of the individual signals strengths, even if there is some degree of correlation between the random propagation effects such as shadowing~\cite{ross2016wireless}, and this Poisson-like behavior is generally more likely for the stronger signals  in the network~\cite{keeler2016stronger}. Beyond being mathematically convenient, recent work has demonstrated that Poisson-based models are able to be fitted to real network data from operators, resulting in network models that adequately incorporate randomness, in both a spatial and temporal sense, from deployed cellular networks~\cite{blaszczyszyn2016spatial,hetnets}. 

It is our aim that the results presented here 
can be tested and used to better understand and control cellular networks with device-to-device  capabilities.  Our results focus on the SIR (in the downlink) coverage probabilities of a device connecting to a base station or another device. We present  expressions for coverage probability when the interference comes from base stations as well as other active devices in the network. We pose a useful optimization problem and then demonstrate it  has a simple solution, which we then investigate qualitatively and numerically. 

We do not aim for complete generality in the current work, but rather a proof of concept, which we believe can be extended in a number of ways. For example, under a Poisson network model, it was shown that it is possible to derive, in certain cases, remarkably  simple expressions for the SIR coverage probability when successive interference cancellation is implemented~\cite{zhang2014performance}\cite[Section V.B.]{sinrmoments}. Alternatively, one could consider multi-tier  models of heterogeneous networks~\cite{mukherjee2012distribution,dhillon2012modeling,madhusudhanan2014downlink} or  other device-to-device models~\cite{lin2014spectrum,elsawy2014analytical}.

\section{Network model}
Let $\Phi=\{X_i\}_{i\geq 1}$ and $\Psi=\{Y_i\}_{i\geq 1}$ be two independent homogeneous Poisson point processes with densities $\lambdaB$ and $\lambdaD$, which respectively model the locations of base stations and  devices in our network model on the plane $\R^2$. For every base station $X_i\in\Phi$, we let $\fade_i$ be a positive random variable representing the general propagation effects such as those from multi-path fading, shadowing or other seemingly random phenomena perturbing the base-station-to-device signal. Similarly, for every device $Y_i\in\Psi$, let   $\fadeD_i$ be an exponential random variable with unit mean representing Rayleigh fading experienced by the device-to-device signal.  We assume that all these random variables are independent and often drop the subscript when talking about the general case. We further assume that  each base station and device transmits respectively with constant powers $\PB$ and $\PD$. Finally, to describe the path loss at the origin for a signal originating from a point $x\in\R^2$, we introduce the function 
\begin{equation}
\pathB(x)=(\kappaB|x|)^{\betaB },
\end{equation}
where the constants $\beta>2$ and $\kappa>0$.

Under this popular network model, 
it has been observed a number of times that the signal strengths experience a type of propagation invariance, meaning they only depend on the random propagation effects 
through one key moment
. More specifically, the signal strengths received  from either base stations or devices at a given location, considered as a point process on the real line, form a Poisson point process, where the intensity measures depend solely on the propagation constants
\begin{equation}
\aB=\frac{\lambdaB\pi\E(\fade^{2/\beta})\PB^{2/\beta}}{\kappaB^2},
\end{equation}
and
\begin{equation}
\aD=\frac{\lambdaD\pi\E(\fadeD^{2/\beta})\PD^{2/\beta}}{\kappaD^2},
\end{equation}
respectively describing signals from base stations and devices; see, for example,~\cite[Lemma 1]{hextopoi-journal} for the precise form of the intensity measures. For exponential $\rayleigh$, the key moment $\E(\rayleigh^{2/\beta}) =\Gamma(1+2/\beta)$, where $\Gamma$ is the gamma function. Although we have included terms for the transmitting powers, we can always assume $\PB=1$ and $\PD=1$, and then lift this assumption by simply multiplying the propagation constants $\aB$ and $\aD$ respectively by $\PB^{2/\beta}$ and $\PD^{2/\beta}$. 

\section{Optimization problem}\label{s.optimization}
At any moment in time, we assume there is a proportion $p$ of devices that are active, which means they are transmitting, and a proportion $1-p$ of devices that are non-active, which means they are potentially receiving signals from base stations or other devices. We further assume that the probability of any device being active or not is independent of all the other devices. This de-centralized approach is equivalent to the simplest Aloha medium access protocol and results in a Poisson process of active devices with density $p\lambdaD$.   

To pose our optimization problem, we need, as a constraint, the SIR at a typical non-active device with respect to the base station with interference from all other base stations and active devices, which is a random variable denoted by $\SIR_{\text{B2D}}(p)$, implying that $\SIR_{\text{B2D}}(0)$ is the base station SIR without interference from any active devices. We are also interested in the SIR with respect to a typical active device at distance $r$ with interference from all base stations and other active devices, which is a random variable denoted by $\SIR_{\text{D2D}}(p)$.  One possible choice for $r$ would be the average distance between devices, which in our Poisson network is $1/(2\sqrt{\lambdaD})$. Another choice would be the average distance between base stations $1/(2\sqrt{\lambdaB})$.  These choices might correspond, respectively,
to {\em intra-} or {\em extra-cell device-to-device connections}.

We let $\thresholdD>0$ be the technology-dependent SIR threshold for the signal reception of (non-active) devices. Finally, we define the device spatial throughput~\footnote{Also called spatial density of successful transmissions in~\cite[Section 16.3]{FnT2}} as 
\begin{equation}
\throughput(p):=p\lambdaD\Prob (\SIR_{\text{D2D}}(p) >\thresholdD) \,.
\end{equation}
We now present our optimization problem:
\begin{equation}\label{e.the-problem}
\begin{aligned}
& \underset{0\leq p \leq 1}{\text{maximize}}
& & \throughput(p) 
\\
& \text{subject to}
& & \inf_{\thresholdB\ge 1}\frac{ \Prob (\SIR_{\text{B2D}}(p)> \thresholdB)}{ \Prob (\SIR_{\text{B2D}}(0)> \thresholdB)}  \ge \delta \,,
\end{aligned}
\end{equation}
where $\thresholdB>0$ is the SIR threshold of base-stations-to-device transmission and $\delta$ ($0\leq\delta\leq1$)  is a parameter, that we call
the \emph{degradation factor}, which quantifies how much of the base station coverage probability is lost, due to additional interference from devices, when devices are allowed to communicate directly with each other.  Note that problem~\eqref{e.the-problem} represents an optimization of the 
spatial throughput of device-to-device  communications, while still maintaining 
the degradation of the distribution of the downlink (base-station-to-device) SIR under control, {\em uniformly} in the domain $\thresholdB\ge1$.

Under our network model, the value of $p$ that maximizes the device spatial throughput, thereby solving problem~\eqref{e.the-problem}, is  given by 
\begin{align} \label{pmaxfirst}
p^*= \min[ 1 , p_1^* , p_2^*] \,,
\end{align}
where
\begin{align}
&p_1^*=\frac{\thresholdD^{-2 / \betaB}}{ \aD r^2 \Gamma(1-2/\betaD) } \label{p1}, \\
&p_2^*=\frac{\aB}{\aD}\left(\frac{1}{\delta}-1\right) \label{p2} \,.
\end{align}
The probability $p_1^*$ (provided it is not larger than one) gives the unconstrained maximum of the device throughput, while $p_2^*$ corresponds to achieving the degradation constraint. In the next sections we present the SIR framework and results that will yield the above expressions for the maximizing value of $p$, but these sections can be skipped to see the numerical results and discussion in Section~\ref{s.discussion}. 

More generally, problem~\eqref{e.the-problem} can be stated and solved  with the  downlink SIR control in the domain
$\thresholdB\ge \thresholdB_{\min}$, for any fixed $\thresholdB_{\min}>0$.
However, it has the above remarkably simple solution only  in the case of the simple coverage domain $\thresholdB\ge \thresholdB_{\min}=1$ 
considered in this short paper.

\section{SIR framework and results}
 We now assume that all devices are active, so $p=1$. We will later remove this assumption by simply replacing the constant $\aD$ (or the device density $\lambdaD$) with the constant $p\aD$ (or the active device density $p\lambdaD$) in all our expressions.

\subsection{Downlink coverage}
Due to stationarity, we can study our model in terms of the downlink SIR at the origin.
In a cellular network model one usually defines the SIR in terms of only signals from the base stations, but in our model we include also interference from active devices, which gives a collection of random variables representing the possible SIR values
\[
\left\{
\frac{\fade_i/\pathB(X_i)}{ \ID  + [\IB - \fade_i /\pathB(X_i) ]  }   :  X_i\in \Phi 
\right\},
\]
where 
\begin{equation}
\IB=\sum_{i=1}^{\infty} [\fade_i/\pathB(X_i)], 
\end{equation}
is the sum of all the base station signals, which we call the total interference, and $\ID$ is an independent random variable representing total interference from all the devices.
  
Without the device interference term $\ID$, the above collection of SIR random variables, which can also be interpreted as a point process on the positive real-line $\R^+$,  has been the basis for most of the stochastic geometry models of wireless networks. As a point process, it has been studied in detail and is a simple function of a much-studied object in mathematics called a Poisson-Dirichlet process or distribution~\cite{sinrPD}. But for our purposes, the largest value of these random variables is of interest, as it is assumed a device connects to the base station that maximizes its SIR value, which means that we are interested in the tail-distribution of the maximum SIR value
\begin{equation}
\calPB(\thresholdB):=\Prob\left(\max_i\left\{
\frac{\fade_i/\pathB(X_i) }{ \ID + [\IB - \fade_i /\pathB(X_i) ]  }    
\right\} > \thresholdB\right)  .
\end{equation}
 The next result can be derived for all $\thresholdB$, but we present it for $\thresholdB\ge 1$, which for $\ID=0$ gives a remarkably simple expression for the coverage probability $\calPB$ 
 ~\cite{dhillon2012modeling}. 
 \begin{lemma}
For $\thresholdB\ge 1$, the probability of a device connecting to a base station, given there is interference from other base stations and devices, is
\begin{align}\label{covprobBlaplace}
\nonumber&\calPB(\thresholdB)\\
&=\frac{2\thresholdB^{-2/\betaB}}{\Gamma(1+2/\betaB)}
\int_0^\infty
\!\!\!\!\!\!u e^{-u^2\Gamma(1-2/\betaB)}\calL_\ID([\aB^{-1/2}
u]^\betaB)\, du\,,
\end{align}
where 
$\calL_{\ID}$ is the Laplace transform of $\ID$. 
When there are no active devices, so $\ID =0$, the above expression reduces to
\begin{equation}\label{covprobBsimple}
\frac{\thresholdB^{-2/\betaB}} {\Gamma(1+2/\betaB) \Gamma(1-2/\betaB)}.
\end{equation} 
\end{lemma}

\begin{IEEEproof}
We conditioned on the device interference $\ID=\nu$, then  the coverage probability is well-known, for example~\cite[equation (58)]{sinrmoments}, and in the regime $\thresholdB\ge 1$,  it is given by
\begin{align*}
\calPB&(\thresholdB| \ID=\nu) \\
=&\frac{2\thresholdB^{-2/\betaB}}{\Gamma(1+2/\betaB)}
\int_0^\infty
\!\!\!\!\!\!u e^{-u^2\Gamma(1-2/\betaB)} e^{ -\nu \aB^{-\betaB/2} u^\betaB }\, du\,.
\end{align*}
But then we remove the conditioning by taking the expectation of $\calPB(\thresholdB| \ID=\nu)$ with respect to $\ID$, yielding
\begin{align*}
\calPB & (\thresholdB)\\ =&\frac{2\threshold^{-2/\betaB}}{\Gamma(1+2/\betaB)}
\E_{\ID}\left[\int_0^\infty
\!\!\!\!\!\!u e^{-u^2\Gamma(1-2/\betaB)} e^{ -\ID \aB^{-\betaB/2} u^\betaB }\, du\, \right]\\
=&\frac{2\thresholdB^{-2/\betaB}}{\Gamma(1+2/\betaB)}
\int_0^\infty
\!\!\!\!\!\!u e^{-u^2\Gamma(1-2/\betaB)} \E_{\ID}\left[e^{ -\ID \aB^{-\betaB/2} u^\betaB }\right]\, du\, .
\end{align*}
But we see that 
$\E_{\ID}[e^{ -\ID \aB^{-\betaB/2} u^\betaB }]$
is the Laplace transform of the random variable $\ID$ with parameter $\aB^{-\betaB/2}
u^\betaB$. 
\end{IEEEproof}

%
%

Interestingly,  the interference term
\begin{equation}
\ID=\sum_{j=1}^{\infty} [\fadeD_j/\pathD(Y_j)] ,
\end{equation}
 is incorporated into the expression for coverage probability $\calPB$  via its Laplace transform,  which, in such an explicit form, we believe is a new observation that we will soon leverage.

\begin{lemma}
The total device interference $\ID$ has the Laplace transform
\begin{equation}\label{laplaceID}
\Lap_{\ID}(\xi )=e^{-\aD \Gamma(1-2/\betaD) \xi^{2/\betaD}}.
\end{equation}

\end{lemma}
\begin{IEEEproof}
For exponential $\fadeD$ with mean $1/\mu$, the Laplace transform of $\ID$ is well-known  (see, for example, \cite[ $2.25$]{FnT1})
\begin{equation}
 \Lap_{\ID}(\xi )=e^{-\lambda(\xi/\mu)^{2/\beta}\pi C(\betaD)/\kappa^2},
\end{equation}
where 
\begin{equation}
 C(\beta)=\Gamma(1-2/\beta)\Gamma(1+2/\beta),
\end{equation}
and where  $\Gamma(1+2/\betaD)$ is the $(2/\betaD)$-moment of an exponential random variable with unit mean.
\end{IEEEproof}



\begin{proposition}
For $\thresholdB\ge 1$, the probability of a device connecting to a base station, given there is interference from other base stations and devices, is
\begin{equation}\label{covprobB}
\calPB(\thresholdB)=\frac{\thresholdB^{-2/\beta}}{\Gamma(1+2/\beta) \Gamma(1-2/\beta) } \frac{\aB}{\aB+ \aD}  \,.
\end{equation}

\end{proposition}

\begin{IEEEproof}
Given the independence between $\IB$ and $\ID$, 
we substitute the Laplace transform (\ref{laplaceID}) into 
 equation~(\ref{covprobBlaplace}), and evaluate the resulting integral 
\begin{align*}
& \int_0^\infty
\!\!\!\!\!\!u e^{-u^2\Gamma(1-2/\betaB)} e^{-\aD \Gamma(1-2/\betaD) (\aB^{-\betaB/2}
u^\betaB)^{2/\betaD}} \, du \\ 
&= \frac{1}{\Gamma(1-2/\beta)(1+\aD /\aB)} \int_0^\infty
\!\!\!\!\!\!u e^{-u^2} \, du 
 \\ 
&= \frac{1}{\Gamma(1-2/\beta)}\frac{\aB}{\aB+ \aD}  \frac{1}{2}\, ,
\end{align*}
which completes the proof.

\end{IEEEproof}

\begin{remark}
The coverage probability expression (\ref{covprobB}) can also be derived from  previous results in the setting of multi-tier networks, such as~\cite[Corollary 3]{dhillon2012modeling}. Furthermore, this probability expression has a intuitive interpretation. The ratio $\aB/(\aB+\aD)$ is simply the probability that the strongest signal belongs to a base station, which can be reasoned via the concept of equivalent networks~\cite[Remark 17]{sinrmoments}. This probability is then multiplied by the coverage probability of the network with both interference from base stations and devices. We could repeat this step by introducing another Poisson process of interfering signals with path loss model $\ell$ and propagation constant, say, $\aI$, and replacing the aforementioned ratio with $\aB/(\aB+\aD+\aI)$. 
\end{remark}

\subsection{Device-to-device coverage}
We now consider the SIR with respect to a typical device, assuming that 
a (hypothetical or virtual~\footnote{The fact that the receiver is not a point of the original device point process  is the simplifying assumption of the bi-polar network model of~\cite{BBM06IT}; see also~\cite[Section~16]{FnT2}.}) receiver is within a distance $r$ to it.
Without loss of generality, we can assume that this receiver is  located at the origin and the typical device, denoted by $Y_0$,  located at the distance $r$ from the origin and independent of the Poisson process of other devices.
The interference received at the origin comes from these
other devices, as well as all base stations. As in the previous section, we denote them respectively by $\ID$ and $\IB$. The  
 Rayleigh fading assumption between the origin and $Y_0$ is represented by an exponential variable $\rayleigh$ with unit mean.
The SIR with respect to the typical device is then represented  by  the random variable
\begin{equation}
\frac{\rayleigh/\pathD(r)} { \IB+ \ID   } ,
\end{equation}
and we define the corresponding coverage probability 
\begin{equation}\label{covprobD}
\calPD(\thresholdD):=\Prob\left[
\frac{\rayleigh/\pathD(r)} {\IB+\ID} >\thresholdD  \right] .
\end{equation}

\begin{proposition}
The probability of the typical device  connecting to a receiver at location~$r$ is equal to 
\begin{equation}
\calPD(\thresholdD)=  e^{- (\aB+\aD) r^2  \Gamma(1-2/\betaB) \thresholdD^{2 / \betaB}}  .
\end{equation}
\end{proposition}

\begin{IEEEproof}
The independence of  $\IB$ and $\ID$ and exponential variable $\rayleigh$ gives 
\begin{align*}
\Prob\left[
\frac{\rayleigh/\pathD(r)}{\IB+\ID} >\thresholdD \right] &
 = \Prob \left[
\rayleigh  >   {\pathD(r)\thresholdD (\IB+\ID )   }  \right]  \\
&=  \calL_{\IB}\left[
\pathD(r)\thresholdD  \right]   \calL_{\ID}\left[\pathD(r)\thresholdD \right]   
\\
&=    e^{- \aB r^2 \Gamma(1-2/\betaB) \thresholdD^{2 / \betaB}}   e^{- \aD  r^2 \Gamma(1-2/\betaD) \thresholdD^{2 / \betaB}},
\end{align*}
where we have used~\eqref{laplaceID} for the Laplace transform of~$\ID$.
\end{IEEEproof}


\subsection{Maximizing device spatial throughput}
We now lift our $p=1$ assumption, by replacing $\aD$ with $p\aD$, and write $\calPB(\thresholdB,p)$ and $\calPD(\thresholdD,p)$ to denote the coverage probabilities (\ref{covprobB}) and (\ref{covprobD}).  
We wish to maximize the spatial throughput $\throughput(p)=p\lambdaD \calPD(\threshold,p)$ by varying $p$, while ensuring the probability of a typical non-active device connecting to a base station corresponds to a certain degradation factor $0\leq \delta \leq 1$. In other words, we seek 
\begin{equation}
p^*=\arg\max_p[p\lambdaD \calPD(\thresholdD,p)],
\end{equation}
with the constraint that for all $\thresholdB\geq1$
\begin{equation}\label{constraint}
\calPB(\thresholdB,p)\ge\delta\calPB(\thresholdB,0) \,.
\end{equation}
We now restate our results from Section~\ref{s.optimization} more formally. 
\begin{theorem}
For our device-to-device communication model, the device spatial throughput  $p\lambdaD\calP(\thresholdD,p)$ with constraint (\ref{constraint}) satisfied for all $\thresholdB\geq1$ is maximized by the value of $p$  given by
$p^*= \min[ 1 , p_1^* , p_2^*] \,,
$
where $p_1^*$ and $p_2*$ are  respectively given by equations (\ref{p1}) and (\ref{p2}).
The device spatial throughput then takes one of three values
\begin{align}
\throughput(1) &= \lambdaD \,  e^{- (\aB +\aD) r^2 \Gamma(1-2/\betaB) \thresholdD^{2 / \betaB}} \label{D1} \\
\throughput(p_1^*) &= \frac{\lambdaD \, e^{- \aB/\aD} e^{-1}} {\aD  r^2 \Gamma(1-2/\betaD) \thresholdD^{2 / \betaB} } \label{Dp1} \\
\throughput(p_2^*) &= \frac{\aB}{\aD}\left(\frac{1}{\delta}-1\right)\lambdaD\, e^{- \aB r^2  \Gamma(1-2/\betaB) \thresholdD^{2 / \betaB}/ \delta }  \label{Dp2}\,,
\end{align}
respectively.
\end{theorem}

\begin{IEEEproof}
We find the (unconstrained) maximum of the device spatial throughput by differentiating 
\begin{equation*}
\nonumber p\lambdaD \calPD(\thresholdD,p) = p\lambdaD \, e^{- (\aB +p\aD ) r^2 \Gamma(1-2/\betaD) \thresholdD^{2 / \betaB}}
\end{equation*}
with respect to $p$ and setting the result to  zero, 
 giving the   value of $p$ that maximizes the device spatial throughput  is $p_1^*$. 
 But degradation constraint (\ref{constraint}), coupled with the coverage probability probability expressions (\ref{covprobBsimple}) and (\ref{covprobB}), implies
 \begin{equation*}
p^*  < \frac{\aB}{\aD}\left(\frac{1}{\delta}-1\right).
\end{equation*}
Explicitly substituting the values of $p$ into the spatial throughput expression completes the proof.
 \end{IEEEproof}

\section{Discussion and numerical results}\label{s.discussion}
For our numerical results, we match the moments of $F$ and $E$, so $\E(\fade^{2/\beta})=\E(\fadeD^{2/\beta})=\Gamma(1+2/\beta)$, but we can use other values for these moments, particularly for $\fade$ whose distribution, we recall, can be very general. For most of the results, we have chosen parameters such that the two propagation constants $\aB$ and $\aD$ are equal to each other, namely $(\PD/\PB)^{2/\beta}=\lambdaD/\lambdaB$, motivated by the scenario where the base stations have larger transmitting powers, but the density of devices is lower. 

\begin{figure}[t!]
\begin{center}
\centerline{\includegraphics[scale=0.5]{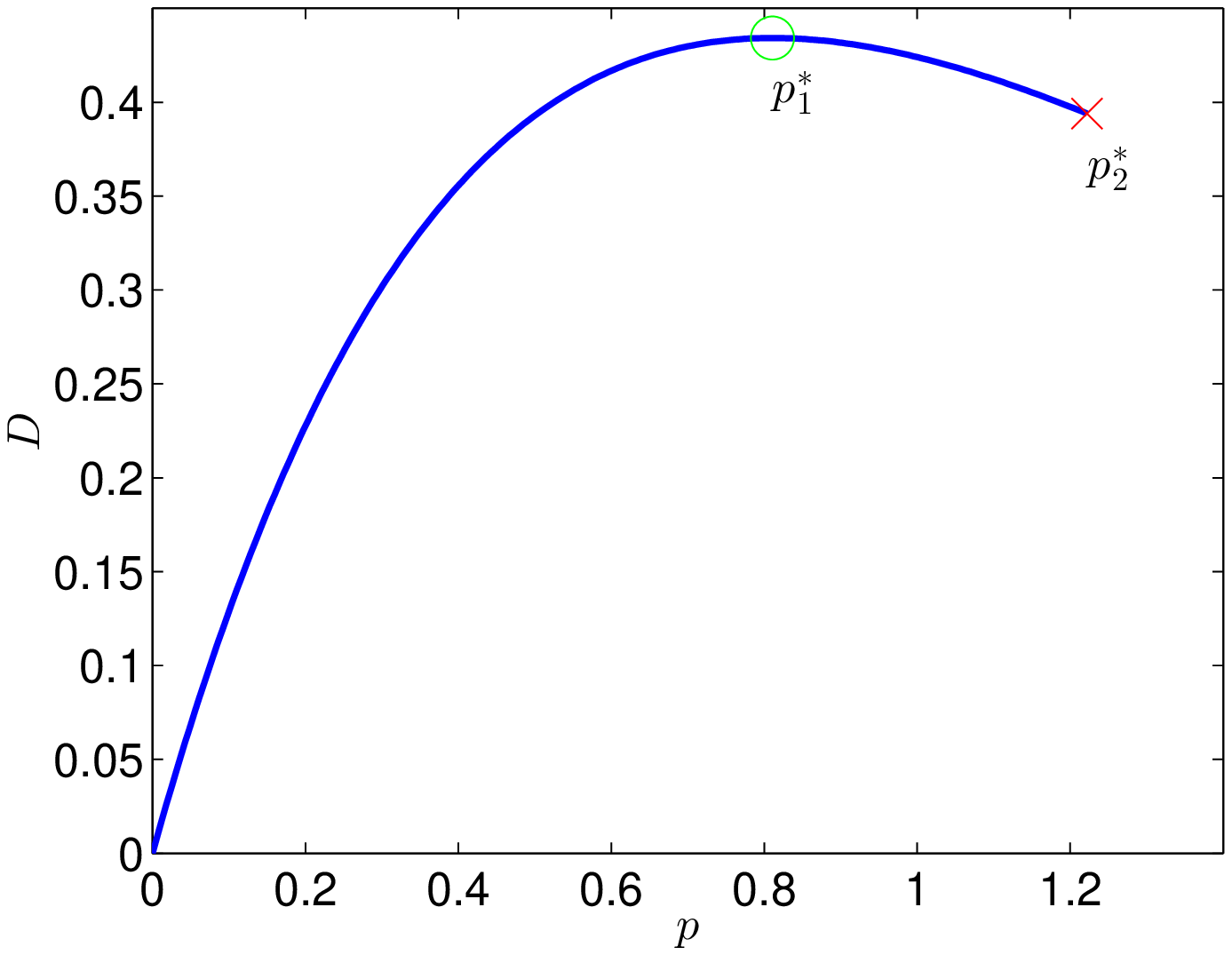}}
\caption{Spatial throughput $D$  with the parameters $\delta=0.45$, $\thresholdD=\thresholdB=1$, $\beta=4$, $\lambdaB=1$, $\lambdaD=5$, $\PB=25$, $\PD=1$, $\E(\fade^{2/\beta})=\E(\fadeD^{2/\beta})=\Gamma(1+2/\beta)$, $\kappa=1$ (so $\aB=\aD$)  and the intra-cell device-to-device case $r=1/(2\sqrt{\lambdaD})$.
\label{Plot1}}
\end{center}
\vspace{-2ex}
\end{figure}

Assuming $p_1^*\leq 1$, there are two principal regimes for values of $p$ that maximizes the spatial throughput. The $p_1^*\leq p_2^*$ case results in the unconstrained optimization solution achieving the maximum, for which an example is shown in Figure~\ref{Plot1} with $r=1/(2\sqrt{\lambdaD})$ (intra-cell device-to-device connections) and the spatial throughput is given by equation~(\ref{Dp1}). Unconstrained optimization is always possible when $p_2^*>1$. Otherwise, to achieve the unconstrained optimum, 
one needs to tolerate the degradation of the downlink coverage probability
\begin{equation}\label{deltastar}
\delta= 
 1-\frac{1}{1+ \aB r^2 \Gamma(1-2/\betaD) \thresholdD^{2 / \betaB}}.
\end{equation}

\begin{figure}[t!]
\begin{center}
\centerline{\includegraphics[scale=0.5]{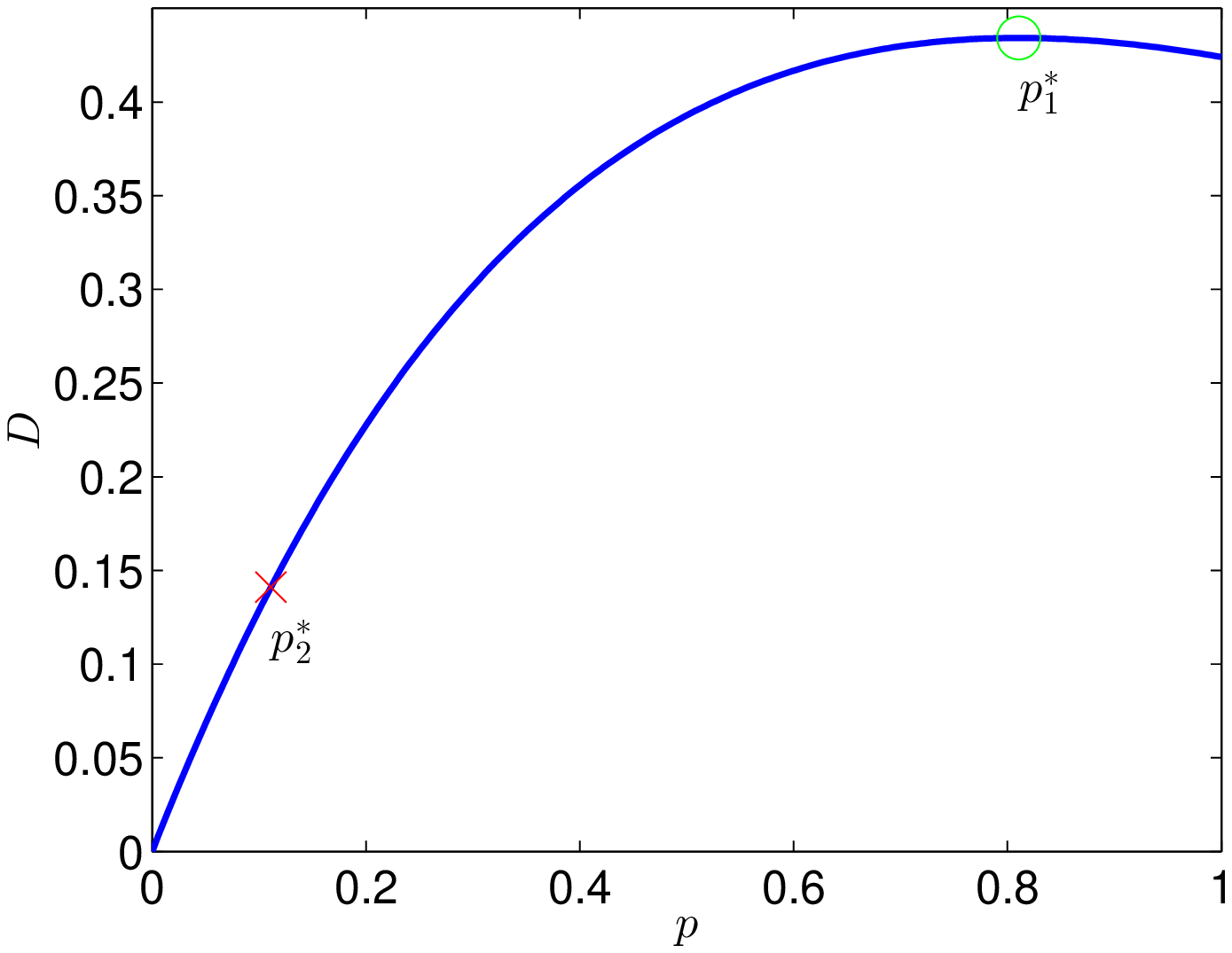}}
\caption{Spatial throughput $D$  with the parameters $\delta=0.9$, $\thresholdD=\thresholdB=1$, $\beta=4$, $\lambdaB=1$, $\lambdaD=5$, $\PB=25$, $\PD=1$, $\E(\fade^{2/\beta})=\E(\fadeD^{2/\beta})=\Gamma(1+2/\beta)$, $\kappa=1$ (so $\aB=\aD$)  and the intra-cell device-to-device case $r=1/(2\sqrt{\lambdaD})$.
\label{Plot3}}
\end{center}
\vspace{-2ex}
\end{figure}

Conversely, when $p_1^* > p_2^*$, the spatial throughput is given by equation~(\ref{Dp2}). If we keep increasing $\delta$ then the degradation constraint (\ref{constraint}) quickly dictates the solution, and we see in Figure~\ref{Plot3} that the majority of devices need to be non-active. This suggests a high cost in terms of base station coverage if we want a reasonably  good spatial throughput
of intra-cell device-to-device communications
(provided that propagation constants $\aB$ and $\aD$ are comparable in magnitude).

\begin{figure}[t!]
\begin{center}
\centerline{\includegraphics[scale=0.5]{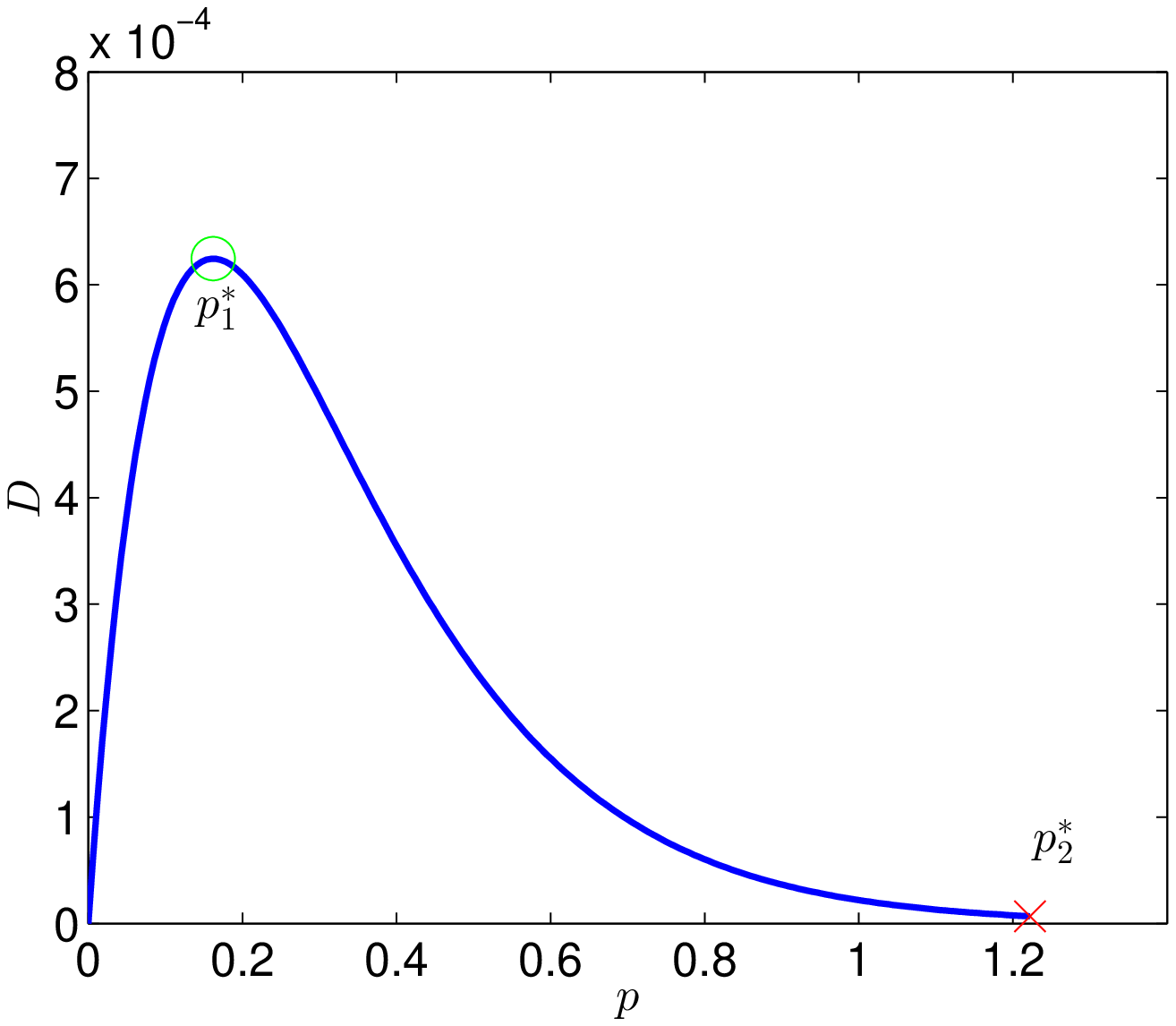}}
\caption{Spatial throughput $D$  with the parameters $\delta=0.45$, $\thresholdD=\thresholdB=1$, $\beta=4$, $\lambdaB=1$, $\lambdaD=5$, $\PB=25$, $\PD=1$, $\E(\fade^{2/\beta})=\E(\fadeD^{2/\beta})=\Gamma(1+2/\beta)$, $\kappa=1$ (so $\aB=\aD$)  and the extra-cell device-to-device case $r=1/(2\sqrt{\lambdaB})$.
\label{Plot4}}
\end{center}
\vspace{-2ex}
\end{figure}

Of course,  the choice of $r$ has a large effect on the results, illustrated in the difference between Figure~\ref{Plot1}, where the intra-cell device-to-device case $r=1/(2\sqrt{\lambdaD})$, and  Figure~\ref{Plot4}, where extra-cell device-to-device case $r=1/(2\sqrt{\lambdaB})$ is considered. In particular, in the case of  extra-cell device-to-device communications,
the downlink coverage constraint has no significant impact on the optimization of the device spatial throughput, being attained at smaller value of $p$, which is anyway much smaller, than in the case of  the intra-cell device-to-device case communications.
Indeed, the choice of $r$ has no effect on $p_2^*$, but 
but $p_1^*$ is decreasing  in $r$.

We can also study effects of device intensity $\lambdaD$ on the spatial throughput. Independently of $r$ 
\begin{equation}
p_2^*=\frac{\lambdaB }{\lambdaD }\frac{\PB^{2 / \betaB}\E(\fade^{2/\beta}) }{\PD^{2 / \betaB} \Gamma(1+2/\betaD) }\left(\frac{1}{\delta}-1\right).
\end{equation}

\begin{figure}[t!]
\begin{center}
\centerline{\includegraphics[scale=0.5]{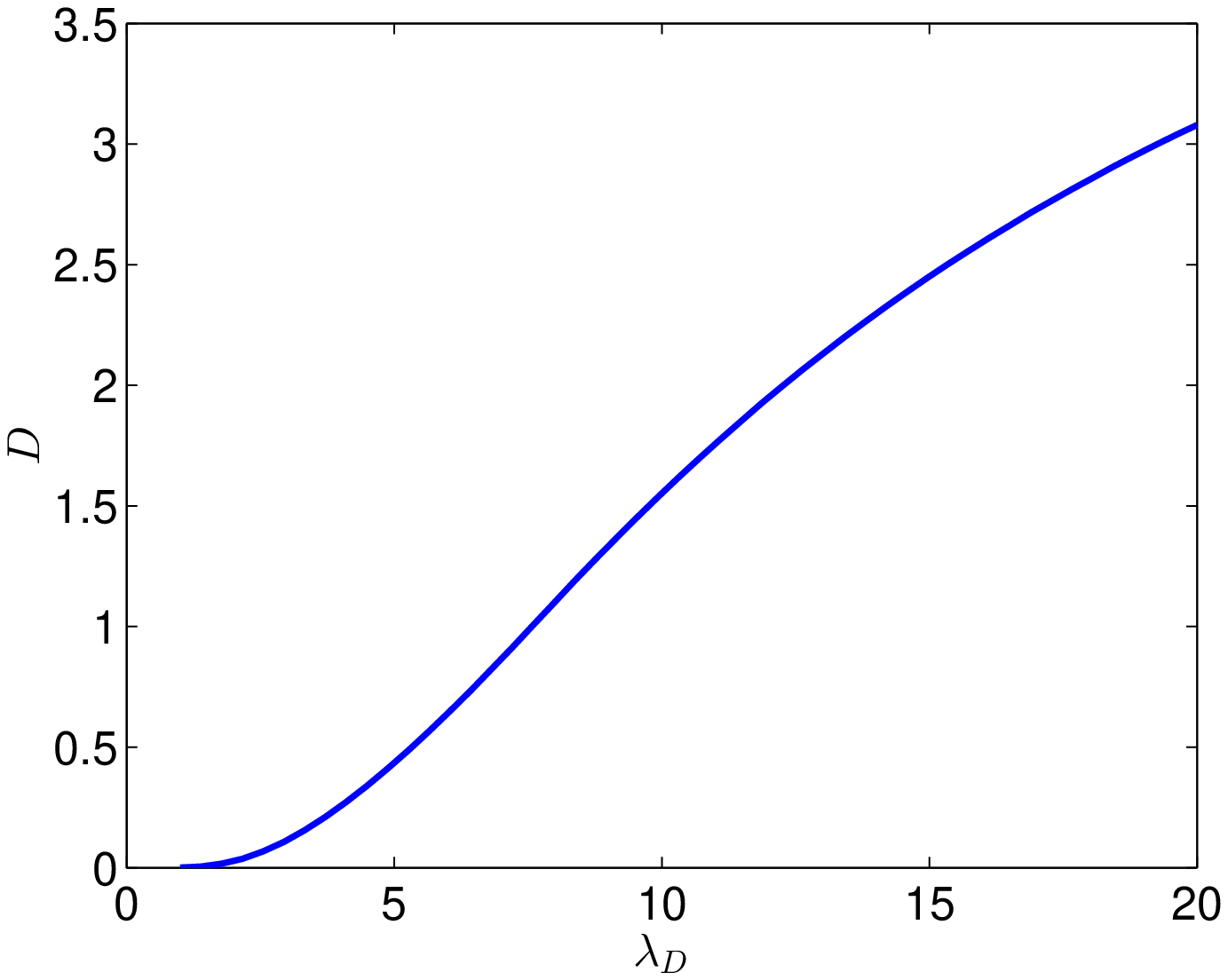}}
\caption{Intra-cell device-to-device case: spatial throughput $D$  with parameters  $r=1/(2\sqrt{\lambdaD})$, $\delta=0.45$, $\thresholdD=\thresholdB=1$, $\beta=4$, $\lambdaB=1$, , $\PB=25$, $\PD=1$, $\E(\fade^{2/\beta})=\E(\fadeD^{2/\beta})=\Gamma(1+2/\beta)$, and $\kappa=1$.
\label{LambdaDD}}
\end{center}
\vspace{-2ex}
\end{figure}

Regarding $p_1^*$, we first we we look at the intra-cell device-to-device case by setting $r=1/(2\sqrt{\lambdaD})$, giving
\begin{equation}
p_1^*=\frac{4\kappa^2\thresholdD^{-2 / \betaB}}{  \pi \PD^{2 / \betaB} \Gamma(1+2/\betaD) \Gamma(1-2/\betaD) } ,
\end{equation}
and we see 
that $p_1$ becomes independent of $\lambdaB$ or $\lambdaD$, while $p_2$ decreases as $\lambdaD$ increases.
Consequently, the device spatial throughput 
has the functional form 
\begin{align}
\throughput(1) &= c_1 \lambdaD e^{- d_1\lambdaB / \lambdaD }    \\
\throughput(p_1^*) &= c_2 \lambdaD e^{- d_2\lambdaB/\lambdaD }    \\
\throughput(p_2^*) & =c_3 \lambdaB e^{- d_3\lambdaB/\lambdaD}  ,
\end{align}
where  $c_1>0$, $c_2>0$, $c_3>0$, $d_1>0$, $d_2>0$  and  $d_3>0$ are not dependent on $\lambdaD$ or $\lambdaB$. (The exact values are of course easily obtained.) We see in Figure~\ref{LambdaDD} that the spatial throughput increases as the device intensity $\lambdaD$ increases. 

For the extra-cell device-to-device case $r=1/(2\sqrt{\lambdaB})$, we have
\begin{equation}
p_1^*=\frac{\lambdaB }{\lambdaD }\frac{4\kappa^2\thresholdD^{-2 / \betaB}}{  \pi \PD^{2 / \betaB} \Gamma(1+2/\betaD) \Gamma(1-2/\betaD) } \,,
\end{equation}
so now
$p_1^*$ and $p_2^*$ have the same dependence on $\lambdaD$ and $\lambdaB$, and the resulting two curves do not intersect. In other words, if $p_1^* \leq p_2^*$ (or $p_1^*> p_2^*$), for a certain set of parameters, then this will remain the case regardless of the value of $\lambdaD$ or $\lambdaB$
. The device spatial throughput now has the  form
\begin{align}
\throughput(1) &= c_4 \lambdaD\, e^{- d_4\lambdaD/ \lambdaB }    \\
\throughput(p_1^*) & =c_5 \lambdaB\, e^{- d_5 \lambdaB /\lambdaD }    \\
\throughput(p_2^*) & =c_6 \lambdaB\, ,
\end{align}
where $c_4>0$, $c_5>0$ , $c_6>0$,  $d_4>0$ and  $d_5>0$. Now the device throughput is either  weakly dependent on $\lambdaD$, when $p_1^* \leq p_2^*$, or completely independent of $\lambdaD$, when  $p_1^*> p_2^*$, not surprisingly, as the dependence on $\lambdaD$ in our choice for $r$ has been removed.




\section{Conclusion}
We presented an optimization formulation for device-to-device networks in terms of SIR and coverage probability. We showed that if we wish to maximize the device spatial throughput of such a network, then the optimal proportion of active devices is given by a simple formula, serving as a useful guide. These results suggest that the price for allowing device-to-device communication is potentially high in terms of base-station-to-device coverage, but further investigation is needed. A key step in our results is introducing the interference of other Poisson networks into our SIR expressions, which gives another way to model heterogeneous networks with closed and open access. These results demonstrate once again the tractability of the Poisson network models. Natural research directions include comparing these analytic results to simulation and experimental work, as well as studying methods and protocols aimed at preserving device-to-base-station coverage.

\section*{Acknowledgements}
Paul Keeler acknowledges the support of the Leibniz program ``Probabilistic methods for mobile ad-hoc networks''. 

{\small \bibliographystyle{IEEEtran}
\bibliography{Device}
}
\end{document}